\documentclass[twocolumn,prb,groupedaddress,showpacs]{revtex4}
\usepackage{epsfig}
\begin{document}
\renewcommand{\baselinestretch}{1}
\title{Electron relaxation in a double quantum dot through two-phonon
processes}
%
%
\author{V.N. Stavrou and Xuedong Hu}
\address{Department of Physics, State University of New York at Buffalo, 
New York 14260, USA}
\begin{abstract}
We theoretically study the relaxation of electron orbital states of a double
quantum dot system due to two-phonon processes.  In particular, we calculate
how the relaxation rates depend on the separation distance between the
quantum dots, the strength of quantum dot confinement, and the lattice
temperature.  Enhancement of the rates by specific inter-dot distances and
lattice temperatures, and the relative strength of different scattering
channels are discussed.  Our results show that although at low temperatures
($T \sim 1$ K) two-phonon processes are almost four orders of magnitude
weaker compared to one-phonon processes in relaxing electron orbital states,
at room temperature they are as important as one-phonon processes.
\end{abstract}
\pacs{73.21.La, 
71.38.-k, 
03.67.Lx, 
85.35.Be 
}
\maketitle
\date{\today}
%
%
%
\section{INTRODUCTION}

Semiconductor and superconductor nanostructure based quantum computing has
attracted wide spread attention in recent years.\cite{Hu}  Specifically,
single electrons trapped in a semiconductor double quantum dot (QD) have been
suggested as candidates of quantum bits (qubits), where the location of the
electron in the two dots represents the two states of the
qubit.\cite{Barenco,Ekert,Sherwin,Tanamoto,Larionov,Hollenberg}  The proposal
of these so-called charge qubits has prompted careful experimental
investigations \cite{Hayashi,Petta} during the past two years, which in turn
motivated several detailed theoretical studies of the decoherence properties
of these charge qubits.\cite{Barrett,Fed,Voro,Stavrou_Hu2005}  So far all
the calculations on electron relaxation due to electron-phonon interaction
have focused on one-phonon processes, specifically single-phonon emission
process.  Here we calculate the relaxation rates due to two-phonon processes
and investigate whether there exist any regime where two-phonon processes
might be as important as one-phonon processes to charge qubit decoherence.

Two-phonon processes in a single three dimensional isotropic GaAs QD
were studied in the context of electron relaxation in photoluminescence
experiments more than a decade ago,\cite{Inoshita92} where processes of the
most interest involve a longitudinal acoustic (LA) and a longitudinal optical
(LO) phonon in the bulk phonon approximation.\cite{Essex}  Here we study the
two-phonon processes in the context of relaxation of double dot charge
qubits, where the qubit energy splitting ($\leq 1$ meV) is much smaller than
that studied before ($\geq 36$ meV, when LO phonons are generally involved)
so that our focus will be on two-acoustic-phonon processes.  In these
processes phonon density of state, qubit energy splitting, and lattice
temperature together lead to interesting behaviors in the overall relaxation
rates and relative strength of the different scattering channels.

The paper is organized as follows.  In Section II we first briefly describe
the electronic states we study and the electron-phonon coupling in our system. 
We then give the expressions for the two-phonon relaxation rates.  In Section
III we present our calculation results, showing how the relaxation rates
depend on double dot parameters such as inter-dot distance and single dot
confinement, and how these rates vary with temperature.  We also discuss the
physical mechanisms behind these obtained behaviors.  In Section IV we draw
some conclusions.

\section{THEORETICAL FORMALISM}

Our model system consists of two coupled QDs separated by a distance of
$2\alpha$, each of which is described by a two-dimensional (2D) harmonic
well.\cite{HD}  The single-dot one-electron wavefunctions are 2D harmonic
oscillator functions \cite{Jacak,Bockel} and are described in terms of the
principal quantum number ${\it{n}}={0,1,2,...}$ and the angular momentum
quantum number ${\it{m}}={0,\pm1,\pm2,...}$ as
\begin {equation}
\label{psi_xy}
\psi_{\parallel}^{(n,m)}\left(\tilde{\rho}, \theta \right) = 
\sqrt{\frac{{n}!}{\pi l^{2}\left({n}+\left|{\it{m}}\right|\right)!}}
\tilde{\rho}^{\left|{\it{m}}\right|}e^{-\tilde{\rho}^{2}/2}
e^{{\it{im}\theta}}
{\mathcal{L}}_{{n}}^{\left|{\it{m}}\right|}\left(\tilde{\rho}^{2}\right)
\end {equation}
where ${\mathcal{L}}_{n}^{\left|{\it{m}}\right|}\left(\tilde{\rho}^{2}\right)$
are the Laguerre polynomials, and $\tilde{\rho} = |{\bf r}_{\parallel}|/{\it
l}$ is a scaled radius, with ${\it l}=\sqrt{\hbar/m^{\ast}\omega_{0}}$.  The
corresponding eigenvalues are
\begin {equation}
E_{nm} = \left(2n+\left|{\it{m}}\right|+1\right)\hbar\omega_{0} \,.
\end {equation} 
Along the growth direction we assume an infinite quantum well (QW)
confinement, 
so that the electron wavefunction has the form 
\begin {equation}
\psi_{\it{z}}\left(\it{z}\right) = \frac{1}{\sqrt{L_z}} cos\left(\pi z/2L_{z}
\right)
\end {equation}
For two QDs that are horizontally coupled, we use a simple in-plane
confinement of two parabolic wells separated by an inter-dot distance
$2\alpha$:
\begin {equation}
\label{Vc}
V_{c} = \frac{1}{2}m^{\ast}\omega_{0}^{2} \ {\rm min} \{
\left(x-\alpha\right)^{2}+y^{2},~\ \left(x+\alpha\right)^{2}+y^{2} \}\,.
\end {equation}
The single electron wavefunction for the lateral direction can in general
be expressed as a superposition of the single-dot wavefunctions: 
\begin {equation}
\label{superposition}
\left |\Psi_{\|}\right> = \sum_{k}{C_{k}\left |\psi_{\|,L}^{k}\right> + 
                              D_{k}\left |\psi_{\|,R}^{k}\right>} \,,
\end {equation}
and the total wavefunction of the system of the coupled QDs is
\begin {equation}
\label{wavefunction}
\Psi(\bf{r}) = \Psi_{\|}\left(\bf{r_{\|}}\right)
                \psi_{\it{z}}\left(\it{z}\right) \,.
\end {equation}
In the present study, the in-plane wavefunctions for the coupled-QD are
calculated numerically by direct diagonalization, using reasonable parameters
of a GaAs QW.

Electrons in GaAs interact with both acoustic and optical phonons.  However,
here we do not consider contributions from the optical phonons to the electron
relaxation due to the small energy splitting in the double dot system we
study.  In this work, we calculate two-phonon relaxation rates caused by both
deformation potential and piezoelectric interactions.  The Hamiltonian that
describes these interactions is given by: 
\begin {equation}
\label{phonons1}
H = \sum_{\bf q} \left( \frac{\hbar}{2 \rho_m V \omega_{\bf q}}
\right)^{1/2} {\mathcal{M}}({\bf q})
\rho({\bf q}) (a_{\bf q}+a_{-\bf q}^{\dagger}) \,,
\end {equation}
where $\rho_m$ is the mass density of the host material, $\omega_{\bf q}$ is
the frequency of the phonon mode with wave vector ${\bf q}$, $V$ is the volume
of the sample, $a_{\bf q}$ and $a_{-\bf q}^\dagger$ are phonon annihilation
and creation operators, and $\rho({\bf q})$ is the electron density operator.
The interaction strength $\mathcal{M}({\bf q})$ is defined by
\begin {equation}
\label{phonons2}
\mathcal{M}({\bf q}) = D\left| \bf{q} \right| + {\it
i}\mathcal{M}_{\lambda}(\hat {q}) \,,
\end {equation}
where the first term represents the deformation potential interaction with
deformation constant $D$, and the second term, which is imaginary, describes
the piezoelectric interaction.  For zincblende crystals (e.g. GaAs), the
piezoelectric term $\mathcal{M}_{\lambda}(\hat {q})$ takes the form
\begin {equation}
\label{PZ}
{\mathcal{M}}^{pz}_{\lambda} (\hat{\bf q}) = 2 e \ e_{14}\left( \hat{q}_{x}
\hat{q}_{y} \xi_{z} + \hat{q}_{y} \hat{q}_{z} \xi_{x} + \hat{q}_{x}
\hat{q}_{z} \xi_{y} \right)
\end {equation}

Using second-order perturbation theory, scattering rates due to the emission
and/or absorption of two LA phonons can be obtained
\begin {eqnarray}
\Gamma_{++}  &=&  \frac{\pi}{\hbar}
\sum_{{\bf q},{\bf k}}
\left| \sum_{s} 
\left( \frac{M_{\bf q}^{is} M_{\bf k}^{sf}}{E_{i}-E_{s}-E_{\bf q}} +
\frac{M_{\bf k}^{is}~M_{\bf q}^{sf}}{E_{i}-E_{s} - E_{\bf k}} \right)
\right|^{2} 
\nonumber \\ 
& & \times \left( N_{\bf q} + 1 \right) \left( N_{\bf k} + 1 \right) 
\delta \left(E_{i} - E_{f} - E_{\bf q} - E_{\bf k} \right) 
\label{emission} \\
\Gamma_{--}  &=&  \frac{\pi}{\hbar}
\sum_{{\bf q},{\bf k}}
\left| \sum_{s} 
\left( \frac{M_{\bf q}^{is} M_{\bf k}^{sf}}{E_{i} - E_{s} + E_{\bf q}} +
\frac{M_{\bf k}^{is}~M_{\bf q}^{sf}}{E_{i} - E_{s} + E_{\bf k}} \right)
\right|^{2} 
\nonumber \\ 
& & \times N_{\bf q} N_{\bf k} 
\ \delta \left(E_{i} - E_{f} + E_{\bf q} + E_{\bf k} \right) 
\label{absorption} \\
\Gamma_{+-}  &=&  \frac{2\pi}{\hbar}
\sum_{{\bf q},{\bf k}}
\left| \sum_{s} 
\left( \frac{M_{\bf q}^{is} M_{\bf k}^{sf}}{E_{i}-E_{s}-E_{\bf q}} 
+ \frac{M_{\bf k}^{is} M_{\bf q}^{sf}}{E_{i}-E_{s} + E_{\bf k}} \right)
\right|^{2} 
\nonumber\\ & &
\times N_{\bf k} \left(N_{\bf q}+1\right) 
\delta\left(E_{i} - E_{f} - E_{\bf q} + E_{\bf k} \right) \,.
\label{emission+absorption}
\end {eqnarray}
Here emission of two phonons (+LA+LA), absorption of two phonons (-LA-LA), and
emission and absorption of one phonon each (+LA-LA, including both overall
emission and absorption) are indicated by subscripts $++$, $--$, and $+-$. 
The later two processes vanish at $T=0$ K.  However, since experimental
temperature is often in the same order as the qubit energy splitting, we
include all two-phonon processes in our calculation.  Indices $i$ and $f$
represent the initial and final electronic states, which are the ground and
first excited double dot states respectively.  Index $s$ refers to
intermediate electronic states which are among the 13 higher energy states we
include in the present calculation (the summation over s excludes the initial
and final states, and the sum has converged).  The matrix elements are
calculated as in Ref.~\onlinecite{Stavrou_Hu2005}.
%
%
Notice here that both deformation potential and piezoelectric interactions are
included in the matrix element---they cannot be treated separately as in
one-phonon processes.  $N_{\bf k}$ is the Bose distribution function for the
${\bf k}$ phonon mode with energy $E_{\bf k} = \hbar \omega_{\bf k}$.  The
integrals over ${\bf k}$ and ${\bf q}$ in Eq.~(\ref{emission}) to
(\ref{emission+absorption}) are calculated by Monte Carlo code.

\section{RESULTS AND DISCUSSIONS}

At low temperatures, which are the operating temperature of the current
generation charge-qubit-based architectures, two-phonon absorption processes
should be much weaker than emission processes.  Therefore we first evaluate
the rates for an electron in the excited qubit state (the first excited state
of the double dot as the initial state) to relax to the ground qubit state
(final state) via two-phonon processes.  Two channels are included here: (a)
emission of two LA phonons (+LA+LA) and (b) emission of a high energy LA
phonon and absorption of a lower energy LA phonon (+LA-LA). Throughout the
paper the QW width is fixed at $2L_{z}=10$ nm and the material parameters are
taken from Bruus {\it et al.}\cite{Bruus1993} 

\begin{figure}[tbp]
\begin{center}
\includegraphics[trim= 0 0 0 -45, width=3.0in]{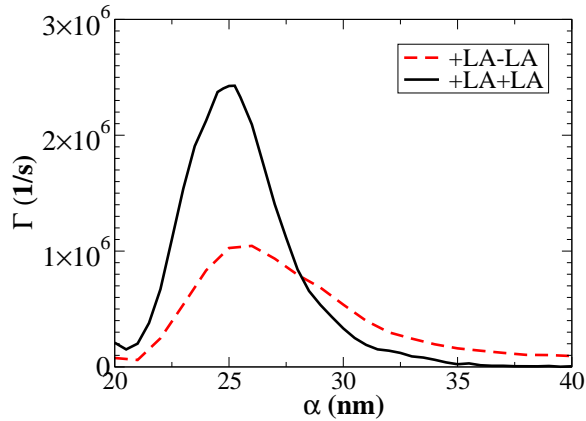}
\protect\caption{Relaxation rates of an electron in a double dot due to two
different two-phonon processes (+LA+LA and +LA-LA) as functions of the half
interdot distance $\alpha$.  The confinement strength is $\hbar\omega=
3$ meV, lattice temperature $T=1$ K, and QW width $2L_{z}=10$ nm.  The
oscillatory behavior of the +LA+LA process at larger $\alpha$ is an artifact
of the integration process.}
\label{fig_1}
\end{center}
\end{figure}
\begin{figure}[tbp]
\begin{center}
\includegraphics[trim= 0 0 0 -45, width=3.0in]{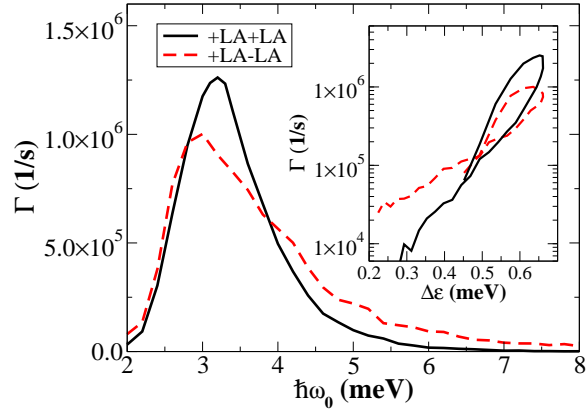}
\protect\caption{Relaxation rates of an electron due to two different 
two-phonon (overall emission) processes (+LA+LA and +LA-LA) as a function of
the confinement strength.  The rates versus the energy splitting $\Delta
\varepsilon$ are shown in the inset.  The half interdot distance is $\alpha=
25$ nm, and the lattice temperature is $T=1$ K.}
\label{fig_2}
\end{center}
\end{figure}

In Figs.~\ref{fig_1} and \ref{fig_2} we present the calculated relaxation
rates due to the two overall emitting processes as functions of half inter-dot
distance $\alpha$ and single dot confinement energy.  For both figures the
lattice temperature is fixed at $T = 1$ K, a usual operating temperature
for semiconductor charge qubits.  The results presented in Fig.~\ref{fig_1}
have some interesting features.  The most prominent is that the order of 
magnitude for both processes is much smaller compared to the one-phonon 
processes\cite{Stavrou_Hu2005}.  Therefore at low temperatures it should be
sufficient to consider only the single-phonon processes when studying 
phonon-induced charge qubit decoherence.  

Both curves in Fig.~\ref{fig_1} display a clear peak at an intermediate
inter-dot distance.  This feature is an interplay between phonon density of
state and the magnitude of the electron-phonon matrix element.  When
inter-dot distance increases, the qubit energy splitting decreases
monotonically.  This leads to a decreasing phonon density of state as it
varies as $\omega^2$ as a function of phonon frequency.  On the other hand,
the magnitude of electron-phonon matrix element increases with increasing
$\alpha$.\cite{Stavrou_Hu2005}  These two opposite trends dictate that a
maximum should develop at an intermediate inter-dot distance for the electron
relaxation rate.  

One additional feature of Fig.~\ref{fig_1} is the crossover between the two
curves.  The physics behind this crossover is even more complicated than the
maxima of the two curves at an intermediate value of $\alpha$.  Qualitatively,
for +LA+LA processes, both phonons involved have to have energies smaller than
the qubit energy splitting, while for +LA-LA processes the emitted phonon has
to have energy larger than the qubit energy splitting.  As bulk acoustic
phonons have a density of state proportional to $\omega^2$, the +LA-LA 
processes should be favored over +LA+LA processes.  On the other hand, 
+LA-LA processes involve absorption of one phonon, which is thermally 
constrained so that its energy has to be relatively small (thus with a 
relatively small phonon density of state) compared to the thermal energy 
(in the current case $\sim 0.1$ meV).  This argument favors the +LA+LA
processes.  Furthermore, we mentioned above that the electron-phonon 
interaction matrix element increases with increasing $\alpha$.  
Therefore, which process yields a higher relaxation rate is determined 
by considering all three factors of lattice temperature, electron-phonon 
matrix element, and phonon density of state.    

The results presented in Fig.~\ref{fig_2} are obtained at a fixed interdot
distance of $2\alpha=50$ nm.  Notice that here the maxima of the two 
relaxation rate curves are mostly determined by the qubit energy splitting
(as shown in the inset) and the phonon density of state consideration.  The
matrix element plays a less important role because it is not as sensitive to
the single-dot confinement energy as it is to inter-dot distance.  Again 
the two rates have a crossover, with +LA+LA processes faster at larger qubit
energy splittings while +LA-LA processes faster for smaller qubit energy
splittings.  The physical mechanism is similar to that in Fig.~\ref{fig_1}.

\begin{figure}[tbp]
\begin{center}
\includegraphics[trim= 0 0 0 -45, width=3.0in]{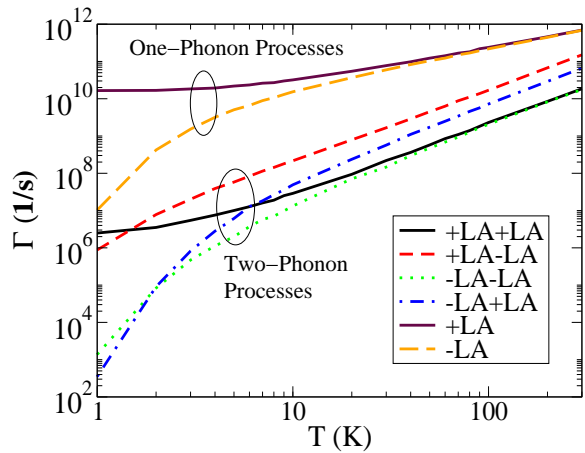}
\protect\caption{Relaxation rates of an electron due to all the two-phonon
processes [+LA+LA, +LA-LA (overall emission), -LA-LA (two-phonon absorption),
and -LA+LA (overall absorption)] and one-phonon processes [+LA (one-phonon
emission) and -LA (one-phonon absorption)] as functions of the lattice
temperature.  Here the confinement strength is $\hbar\omega= 3$ meV, 
and the half interdot distance is $\alpha = 25~meV$.}
\label{fig_3}
\end{center}
\end{figure}

In Fig.~\ref{fig_3} we present the electron relaxation rates as functions
of the lattice temperature for all the one-phonon and two-phonon processes.
At $T = 1$ K, the strongest two two-phonon processes are the +LA+LA and +LA-LA
overall emission process, but the corresponding relaxation rates are more
than three orders of magnitude smaller than the one-phonon emission process. 
The other two two-phonon processes are thermally suppressed because of the
involvement of absorption of high energy phonon(s).  On the other hand, at
room temperature (corresponding to a thermal energy in the order of 30 meV,
much larger than the qubit energy splitting of $< 1$ meV), the strongest
two-phonon process, the +LA-LA (overall emission) process is only one order
of magnitude smaller than the one-phonon processes (both emission and
absorption).  In other words, if the charge qubit states of a double quantum
dot is to be used for some form of quantum information processing at high
temperatures such as room temperature, two-phonon relaxation processes should
also be taken into consideration.

At intermediate temperatures the relaxation rates have crossovers between the
one-phonon-emission-one-phonon-absorption processes and the two-phonon
absorption or emission processes.  The physical mechanism is similar to what
we have discussed for Figs.~\ref{fig_1} and \ref{fig_2}, in that the thermal
suppression of the higher energy phonons (which are only involved in +LA-LA
and -LA+LA processes and have higher density of state) is gradually removed
as temperature increases.  At room temperature the +LA+LA and -LA-LA
processes have almost the same relaxation rates because the thermal factors
in Eqs.~(\ref{emission}) and (\ref{absorption}) are approximately the same
for phonons with energy smaller than 1 meV (recall that for two-phonon
emission or absorption processes each of the phonons involved should have
energy smaller than qubit energy splitting).  Similarly, +LA-LA (overall
emission) and -LA+LA (overall absorption) processes also have similar rates
at high temperatures.  The discrepancy is mostly due to the involvement of
higher energy phonons in these processes, for which the factors $N_k$ and
$(N_k + 1)$ could be sufficiently different.

\section{CONCLUSIONS}

In this study we have investigated two-acoustic-phonon processes (including
both phonon emissions and absorptions) induced charge qubit relaxation in a
semiconductor double quantum dot.  Our results show interesting dependence of
the relaxation rates on the lattice temperature and system configuration
parameters such as inter-dot distance and single dot confinement energy.  We
have found that acoustic phonon density of state, electron-phonon coupling
matrix element, and phonon thermal distributions together lead to crossovers
between different two-phonon scattering channels.  In the context of charge
qubit based quantum information processing, the two-phonon processes are much
weaker than the one-phonon processes at the relevant low temperatures,
although their magnitudes are similar at the room temperature.

\section{ACKNOWLEDGMENT}
%
The work is supported in part by NSA and ARDA under ARO contract
No.~DAAD19-03-1-0128.
%

%
%
%

\begin{thebibliography}{10}
%
%
\bibitem{Hu} X. Hu and S. Das Sarma, Phys. Stat. Sol. (b) {\bf 238}, 260
(2003); X. Hu, cond-mat/0411012, a brief review on quantum dot quantum
computing. 
%
\bibitem{Barenco} A. Barenco, D. Deutsch, A. Ekert, and R. Jozsa, Phys. Rev.
Lett. {\bf 74}, 4083 (1995).
%
\bibitem{Ekert} A. Ekert and R. Josza, Rev. Mod. Phys. {\bf 68}, 733 (1996).
%
\bibitem{Sherwin} M. S. Sherwin, A. Imamoglu, and T. Montroy, Phys. Rev. A
{\bf 60}, 3508 (1999).
%
\bibitem{Tanamoto} T. Tanamoto, Phys. Rev. A {\bf 61}, 022305 (2000).
%
\bibitem{Larionov} A.A. Larionov, L.E. Fedichkin, and K.A. Valiev,
Nanotechnology {\bf 12}, 536 (2001).
%
\bibitem{Hollenberg} L.C.L. Hollenberg, A.S. Dzurak, C. Wellard, A.R.
Hamilton, D.J. Reilly, G. J. Milburn, and R.G. Clark, Phys. Rev. B {\bf 69},
113301 (2004).
%
\bibitem{Hayashi} T. Hayashi, T. Fujisawa, H.D. Cheong, Y.H. Jeong, and Y.
Hirayama, Phys. Rev. Lett. {\bf 91}, 226804 (2003).
%
\bibitem{Petta} J.R. Petta, A.C. Johnson, C.M. Marcus, M.P. Hanson, and A. C.
Gossard, Phys. Rev. Lett. {\bf 93}, 186802 (2004).
%
\bibitem{Barrett} S.D. Barrett and G.J. Milburn, Phys. Rev. B {\bf 68}, 155307
(2003).
%
\bibitem{Fed} L. Fedichkin and A. Fedorov, Phys. Rev. A {\bf 69}, 032311
(2004).
%
\bibitem{Voro} S. Vorojtsov, E.R. Mucciolo, and H.U. Baranger, Phys. Rev. B
{\bf 71}, 205322 (2005). 
%
\bibitem{Stavrou_Hu2005} V.N. Stavrou and X. Hu, cond-mat/0503481.  Phys. Rev.
B {\bf 72} (in press).
%
%
\bibitem{Inoshita92}
T. Inoshita and H. Sakaki, Phys.Rev. B {\bf 46}, 7260 (1992). 
%
\bibitem{Essex}
B. K. Ridley, {\em Electrons and Phonons in Semiconductor} (Cambridge
University Press, 1996);
N. C. Constantinou, J. Phys.: Condens. Mattter {\bf 3}, 6859 (1991);
V. N. Stavrou, M Babiker, and C. R. Bennett J. Phys.: Condens. Matter {\bf
13}, 6489 (2001);
V. N. Stavrou, C R Bennett, O.M.M. Al-Dossary, and M Babiker, Phys. Rev. B
{\bf
63}, 205304 (2001).
V. N. Stavrou, Physica B-Condensed Matter {\bf 337}, 87 (2003).
%
\bibitem{HD} X. Hu and S. Das Sarma, Phys. Rev. A {\bf 61}, 062301 (2000).
%
\bibitem{Jacak}
L. Jacak, P. Hawrylak, and A. W\'{o}js, {\em Quantum Dots} (Springer, 1998).
%
\bibitem{Bockel}
U. Bockelmann Phys. Rev. B {\bf 50}, 17271 (1994).
%
\bibitem{Mahan}
Gerald D. Mahan, {\em Many-Particle Physics} (Plenum Press, New York,
1990).
%
\bibitem{Ridley82}
B. K. Ridley, {\em Quantum processes in semiconductors} (Clarendon Press,
Oxford, 1982).
%
\bibitem{MahanPAP1}
G. D. Mahan, in {\em Polarons in Ionic Crystals and Polar Semiconductors}, ed.
J. T. Devreese (North-Holland, Amsterdam, 1972), pp.553-657.
%
\bibitem{Bruus1993}
H. Bruus, K. Flensberg, and H. Smith Phys. Rev. B {\bf 48}, 11144 (1993).
%
\end{thebibliography}
\end{document}